\documentclass{article}
\usepackage[utf8]{inputenc}
\usepackage[T2A]{fontenc}
\usepackage[english]{babel}
\inputencoding{utf8}

\usepackage{amsmath, amssymb, amsthm, nicefrac, mathtools, url}

\usepackage[
    pdfpagemode=UseNone,
    bookmarks=true,
    bookmarksopen=true,
    pdfpagelayout=SinglePage,
    pdfstartview={XYZ null null 1},
    pdfhighlight=/P,
    colorlinks=true,
    linkcolor=blue,
    citecolor=blue,
    urlcolor=blue
    ]{hyperref}
\usepackage{blkarray}
\usepackage{float}

\advance\textwidth 20mm  \advance\oddsidemargin -10mm
\advance\textheight 20mm \advance\topmargin -15mm

\theoremstyle{plain}
\newtheorem{theorem}{Theorem}
\newtheorem{lemma}[theorem]{Lemma}
\newtheorem{proposition}{Proposition}

\theoremstyle{definition}

\def\C{{\mathbb{ C}}}

\mathtoolsset{showonlyrefs,showmanualtags}

\title{{\Large \bf Novikov equations for commuting differential operators of orders 3,4,5}}

\date{ }

\author{G. B. Shabat \thanks{Russian State University for the Humanities, Moscow, Russia. 
E-mail: george.shabat@gmail.com},\, V. V. Sokolov\thanks{Higher School of Modern Mathematics MIPT,  Moscow,   Russia.  E-mail: vsokolov@landau.ac.ru},\, A. V. Tsiganov \thanks{St.Petersburg State University, St.Petersburg, Russia. E-mail: andrey.tsiganov@gmail.com } }

\usepackage{tikz}

\begin{document}
\maketitle

\begin{abstract}
We consider Novikov equations for commutative ring generated by differential operators of orders 3,4,5. We present an explicit Hamiltonian form of these equations. Using the method of compatible Poisson brackets, we find a   separation of variables on a hyperelliptic curve of genus 2 for the Novikov equations. 
\medskip

\end{abstract}


\section{Introduction}
 \setcounter{equation}{0}
 
The problem of commuting ordinary differential operators is a bridge between the theory of integrable systems and algebraic geometry. On the one hand, two commuting operators are connected by the equation of an algebraic curve. On the other hand, partial solutions of integrable models  of solitonic type satisfy a system of ODEs equivalent to commutation conditions. The latter system is often referred to as Novikov equations.

If we are dealing with commuting operators, then it is natural to consider the commutative ring generated by them.  The largest common divisor of the orders of all operators from the ring is called its rank. The theory of rings of rank 1 is the most developed \cite{krich1, drinf1}.

The rings  generated by operators of orders 2 and $2 g+1$ are of rank 1. The corresponding algebraic curve is hyperelliptic of genus $g$. The algebraic solutions of the KdV equation satisfy the commutativity condition for such operators. 

The simplest ring of rank 1, that does not contain an operator of second order, is generated by commuting operators of orders 3, 4 and 5 \cite{krich2, shab1}.
Although this case is theoretically covered  by the general theory, explicit formulas are missing not only for the coefficients of the operators, but also for the Novikov equations themselves. 

I. Krichever's approach is based on the use of the analytical properties of the Baker-Akhiezer function and leads to a description of the commuting operators in terms of $\theta$ or $\sigma$-functions \cite{buch} on the corresponding algebraic curve. 

We use an alternative way based on a detailed study of Novikov's equations.
In Section 2 we bring the overdetermined system of PDEs equivalent to the commutativity conditions to a passive form. As a result, two different cases arise. We 
call them {\bf genus 2 case} and {\bf genus 1 case}\footnote{see the explanations below.}. The existence of two components of the variety, parametrizing the solutions of the Novikov system, indicates that there exists a component on the boundary of the corresponding moduli space. Its algebro-geometric meaning will be clarifyed in the subsequent publications.

For the genus 2 case, the general solution depends on 7 parameters whereas in the genus 1 case the number of parameters is 6. In the section 2.3,  
we integrate the Novikov system, corresponding to the genus 1 case, in elliptic functions.

The 
remaining part of the article is devoted to the Novikov equations in the genus 2 case. In Section 3, we find a linear Hamiltonian structure for them and  Darboux coordinates such that the Novikov system becomes a new integrable case with
 two degrees of freedom having a quadratic Hamiltonian and an additional first integral quartic in momenta. Following the ideas of papers \cite{g-ts,ts11,ts12}, we find another Darboux coordinates in which the Hamiltonian and the quartic integral are reduced to the St\"{a}ckel form. Using the standard St\"{a}ckel theory,  we construct for the Novikov equations a separation of variables  on a hyperelliptic curve of genus two. The general solution for these equations is given in terms of integrals with a variable upper limit.

For the genus 2 case with some effort, following Krichever's general approach, it is possible to express the coefficients of the $L$ operator in terms of $\theta$-functions on the algebraic curve in $\C^3,$  which relates the operators $L, M$ and $N$. At the end of the paper we verify that this spectral curve is birationally equivalent to the separation curve we have found.

\section{Reduction to passive form}
 \setcounter{equation}{0}

  Let us consider the problem of commutation of the operator
$$
L=D^3+ p D+q, \qquad D=\frac{d}{dx}
$$
with operators $M$ and $N$ of orders 4 and 5. A simple computation shows that  these operators have the form
$$
M=D^4 + \Big(\frac{4}{3} p + m_1\Big) D^2+\Big(\frac{2}{3} p_1+ \frac{4}{3} q + m_2\Big) D
+$$$$ \frac{2}{9} p_2+\frac{2}{3} q_1+\frac{2}{9} p^2+\frac{2}{3} m_1 p+m_3,
$$
and
$$
N=D^5+\frac{5}{3} p D^3+ \Big(\frac{5}{3} p_1+ \frac{5}{3} q + n_1\Big) D^2+ $$$$
\Big(\frac{10}{9} p_2+\frac{5}{3} q_1+\frac{5}{9} p^2+ n_2\Big) D+\frac{10}{9} q_2+\frac{10}{9} p q+\frac{2}{3} n_1 p + n_3.
$$
Here
$$
 p=p(x), \qquad p_i=D^i(p)\qquad q=q(x),\qquad q_i=D^i(q),
$$
and $m_i,$ $n_i$ are constants. 
It is convenient to introduce the function $\displaystyle u=q-\frac{p_1}{2}$ instead of $q$.

The commutativity conditions of $[L, M]=0$ consist of two PDEs for $p$ and $u$. Two more equations are provided by the conditions $[L,N]=0$.  The resulting system is given by:

\begin{equation}\label{eq1}
\Big(m_2 p + 2 m_1 u + \frac{4}{3} p u + \frac{2}{3} u_2\Big)_x=0,
\end{equation}

\begin{equation}\label{eq2}
\begin{array}{c}
\displaystyle \Big(-\frac{1}{3} m_1 p^2+m_2 u+\frac{1}{2} m_2 p_1+m_1 u_1-\frac{1}{6} m_1 p_2+\frac{2}{3} u^2 - \\[3mm]
\displaystyle \frac{4}{27} p^3+\frac{2}{3} p u_1+\frac{2}{3} p_1 u-\frac{1}{6} p_1^2-\frac{1}{3}p p_2+\frac{1}{3} u_3-\frac{1}{18} p_4\Big)_x=0,
\end{array}
\end{equation}

\medskip

\begin{equation}\label{eq3}
\displaystyle \Big(n_2 p+2 n_1 u - \frac{5}{27} p^3+\frac{5}{3} u^2 - \frac{5}{12} p_1^2- \frac{5}{9} p p_2 - \frac{1}{9} p_4\Big)_x = 0,
\end{equation}

\begin{equation}
\begin{array}{c}\label{eq4}
 \displaystyle \Big(-\frac{1}{3} n_1 p^2+n_2 u+\frac{1}{2} n_2 p_1+n_1 u_1 - \frac{1}{6} n_1 p_2 - \frac{5}{9} p^2 u -  \\[3mm]
 \displaystyle 
\frac{5}{18} p^2 p_1 +  \frac{5}{3} u u_1-\frac{5}{18} p_1 u_1  -\frac{5}{18} p_2 u  -\frac{25}{36} p_1 p_2  -  \\[3mm]
 \displaystyle\frac{5}{9} p  u_2 
 -\frac{5}{18} p p_3 - \frac{1}{9} u_4 - \frac{1}{18} p_5\Big)_x = 0.
\end{array}
\end{equation}

It is assumed that the function $p$ is non-constant. Otherwise, it would follow from the equation \eqref{eq3} that $u$ is also constant. Then all the differential operators would have constant coefficients and obviously commute with each other.

At the first glance, it is unclear why this system has at least some non-constant solutions.  However, Krichever's general theory predicts the existence of a seven-parameter family of solutions.

\subsection{ Two Cases.}

\medskip

\qquad In the process of bringing the system   \eqref{eq1},\eqref{eq2},\eqref{eq3},\eqref{eq4} to a passive form, all derivatives of $p$ and $u$ of orders $\ge 3$ can be easily expressed through the derivarives of orders $\le 2$.  Substituting these expressions to the system and its differential consequenses, we obtain several differential relations of order 2.  If we eliminate $u''$ from them by the method of resultants, one of such  obtained equations can be factorised:
\begin{equation}\label{components} 
\begin{array}{c}
\Big(-2 t_2 P'^2 + 
   4 t_1 P P'^2 - 2 P^2 P'^2 + 
   12 U U' P' - 
   24 U'^2 P - 
   P'^2 P''\Big)
\times \\[3mm] 	
	\displaystyle\Big(t_2 P + t_1 P^2 - P^3 - 3 U^2 + 
   \frac{3}{4} P'^2 - 
   P P''\Big)=0.
	\end{array}
\end{equation}
Here and below we use the notation 
$$p=P-3 m_1, \qquad u=U-3 m_3, \qquad t_1= 3 m_1, \quad t_2 =  9 (m_1^2 - n_2), \quad t_3 =  -5 m_2 + 4 n_1.$$
The relation \eqref{components}  is the result of rather tedious calculations, which hopefully  will be presented in our further publications. 

\qquad It can be verified that if the second factor is equal to zero, then the passive form is
\begin{equation} \label{DynSys1} 
\begin{array}{l}
\displaystyle P''=t_2 + t_1 P - P^2 - \frac{3 U^2}{P} + \frac{3 P'^2}{
 4 P}, \\[4mm]
\displaystyle U'' = t_3 P +\frac{1}{2} t_1 U+t_2 \frac{U}{2 P} - \frac{3}{2} P U
 +  \frac{U^3}{2 P^2} -\frac{U P'^2}{8 P^2} +\frac{P' U'}{2 P}.
\end{array}
\end{equation}
Equations  \eqref{eq1}-\eqref{eq4} are satisfied in virtue of this dynamical system. Thus, the general solution depends on 7 parameters $t_1,t_2,t_3, P(0),U(0),P'(0),U'(0).$ One of the parameters can be normalized by a consistent scaling of $x, U, P$ and parameters $t_i$. We call this branch the genus 2 case.

The 
genus 1 case corresponds to vanishing of the first factor in \eqref{components}. In this case we have the following passive form:

\begin{equation} \label{DynSys2} 
\begin{array}{l}
\displaystyle  P''=-2 t_2 + 4 t_1 P - 2 P^2 + 
  \frac{12 U U'}{P'} - \frac{24 P U'^2}{P'^2},\\[4mm]
	
\displaystyle 	U'' = t_3 P -\frac{4}{3} P U-2 t_2 \frac{U'}{P'} + 2 t_1 \frac{P U'}{P'} -
   \frac{2 P^2 U'}{3 P'} -\frac{12 U U'^2}{P'^2} -\frac{16 P U'^3}{P^3}.
\end{array}
\end{equation}
Unlike \eqref{DynSys1}, the initial date for this system are not arbitrary but should lie on the following surface:
\begin{equation} \label{Surface}
3 t_3 P'^3 - 4 U P'^3 - 
 6 t_1 P'^2 U' + 
 4 P P'^2 U' + 
 24 U'^3=0.
\end{equation}
It is easy to verify that the vector field corresponding to \eqref{DynSys2} is tangent to this surface. The system \eqref{eq1} - \eqref{eq4} follows from \eqref{DynSys2}, \eqref{Surface}. Since $U$ can be found from \eqref{Surface}, the general solution depends on 6 parameters $t_i$, 
$P(0), P'(0), U'(0)$. As in the previous case, one of the parameters can be normalized by a scaling.

\subsection{Particular solution in the genus 2 case}

Consider the case $U=0$. It follows from \eqref{DynSys1} that $t_3=0$ and the function $P$ satisfies the equation 
$$
P''=\frac{3 P'^2}{4 P}-P^2+t_1 P + t_2.
$$
This equation has the first integral
$$
P^{-\frac{3}{2}} \Big(P'^2+ \frac{4}{3} P^3-4 t_1 P^2+4 t_2 P \Big)+c=0.
$$
If $c=0$ then $P$ is the Weierstrass $\wp$-function\footnote{Up to some easy rescaling.}. In this case there exists a second order differential operator that commute with $L$ and we come to the simplest commutative ring related to the KdV equation.

Otherwise, $P$ satisfies the equation 
$$
P'^2+ \frac{4}{3} P^3-4 t_1 P^2+4 t_2 P+\frac{4}{3} t_3 +c P^{\frac{3}{2}}=0.
$$
Clearly, its solution can be written as the integral with the variable upper limit.

\subsection{General solution in the genus 1 case}

Using \eqref{DynSys2} and \eqref{Surface}, it is easy to verify that $\displaystyle\frac{U_x}{P_x}$ is an integral of motion and therefore
\begin{equation} \label{qu1}
U=k_1 P+k_0.
\end{equation}
Then \eqref{Surface} gives rise to
\begin{equation} \label{qu2}
4 k_0 - 24 k_1^3 + 6 k_1 t_1 - 3 t_3 = 0
\end{equation}
and \eqref{DynSys2} is equivalent to 
\begin{equation} \label{qu3}
P'' = 12 k_0 k_1 - 2 t_2 - 12 k_1^2 P + 4 t_1 P - 2 P^2.
\end{equation}
Thus, $P(x)$ is the Weierstrass $\wp$-function  and the general solution of \eqref{eq1}-\eqref{eq4} is described by \eqref{qu1},\eqref{qu2}, \eqref{qu3}.

\section{Compatible Poisson brackets and separation of variables for 
the genus 2 case }
 
In the genus 2 case the formulas \eqref{eq1}-\eqref{eq4} define first integrals for the system \eqref{DynSys1}. The following two of them are functionally independent:
\begin{equation}\label{I1}
H_1=-t_3 P - t_1 U - \frac{t_2 U}{P} + P U + \frac{U^3}{P^2} - \frac{U P'^2}{4 P^2} + \frac{P' U'}{P},
\end{equation}
and
\begin{equation}\label{I2}
\begin{array}{c}
H_2=\displaystyle-\frac{t_2^2}{4 P} - \frac{1}{4} t_1^2 P - \frac{1}{6} t_2 P + \frac{1}{6} t_1 P^2 - 
 \frac{P^3}{36} - 2 t_3 U + 
 \frac{3 U^2}{2} + \frac{t_2 U^2}{2 P^2} - \\[4mm]
\displaystyle \qquad \frac{t_1 U^2}{2 P} - \frac{U^4}{4 P^3} - 
 \frac{1}{24} P'^2 - \frac{t_2 P'^2}{
 8 P^2} +  
\displaystyle \frac{t_1 P'^2}{8 P} + \frac{
 U^2 P'^2}{8 P^3} - \frac{P'^4}{
 64 P^3} + \frac{U'^2}{P}.
\end{array}
\end{equation}
 The system \eqref{DynSys1} turns out to be Hamiltonian with a linear Poisson bracket whose Poisson tensor in the variables $\{U, U_1=U', P, P_1=P'\}$ has the form
\begin{equation} \label{p-lin}
\left( \begin{array}{cccc}
  0 & \frac{U}{2} & 0 & P \\
  -\frac{U}{2} & 0 & -P & -\frac{P_1}{2} \\
 0 & P & 0 & 0 \\
-P & \frac{P_1}{2} & 0 & 0
\end{array}%
\right).
\end{equation}
The Hamiltonian of \eqref{DynSys1} is equal to $-H_1$. The first integrals $H_1$ and $H_2$ commute with each other with respect to this bracket. The dynamical system which corresponds to  Hamiltonian $H_2$ is the restriction to the variables $P,P_1,U,U_1$ of the Boussinesq system 
$$
p_t = 2 u_1, \qquad  u_t=-\frac{1}{6}\big(p_3+4 p p_1\big),
$$
which has the Lax representation
$$
L_t = \Big[D^2+\frac{2}{3} p, \, L \Big].
$$

It is easy to verify that
\begin{equation}\label{darbu}
q_1=\sqrt{P}, \qquad q_2=\frac{U}{\sqrt{P}}, \qquad p_1=\frac{2 U_1}{\sqrt{P}}, \qquad p_2=\frac{P_1}{\sqrt{P}}.
\end{equation}
are Darboux coordinates for the linear Poisson bracket above, that is
\[   
\{q_1,p_1\}=1\,,\quad \{q_2,p_2\}=1\,,\qquad \{p_1,p_2\}=\{q_1,q_2\}=\{p_1,q_2\}=\{p_2,q_1\}=0\,.
\]
In these coordinates, the Poisson tensor \eqref{p-lin} has the standard form
\begin{equation}\label{p-dar}
\pi=\left( \begin{array}{cccc}
  0 & 0 & 1 & 0 \\
  0& 0 & 0 & 1 \\
 -1 & 0 & 0 & 0 \\
0 & -1 & 0 & 0
\end{array}%
\right)\,.
\end{equation}
The Hamiltonian \eqref{I1} in the new variables is the following polynomial of second degree in momenta $p_i$: 
\begin{equation}\label{H1}
H_1=\frac{1}{2} p_1 p_2 -\frac{q_2}{4 q_1}p_2^2+q_1^3 q_2 + 
   \frac{q_2^3}{q_1} - t_1 q_1 q_2 - t_2 \frac{q_2}{q_1} - t_3 q_1^2
\end{equation}
while $H_2$ \eqref{I2} is the polynomial of the fourth degree:
\begin{equation}\label{H2}
H_2=\Big(\frac{1}{8 q_1} p_2^2 + \frac{3 t_2 - 3 t_1 q_1^2 + q_1^4 - 3 q_2^2}{
    6 q_1}\Big)^2 + \Big(\frac{t_1 t_2}{2} - \frac{1}{4} p_1^2 + 2 t_3 q_1 q_2 -\frac{4}{3}q_1^2q_2^2\Big).
\end{equation}
Thus, \eqref{DynSys1} can be rewritten as the system of Hamiltonian equations
\begin{equation} \label{eq-ham}
\dot{q}_i=-\frac{\partial H_1}{\partial p_i}\,,\qquad \dot{p}_i=\frac{\partial H_1}{\partial q_i}\,,\qquad i=1,2,
\end{equation}
with two independent first integrals in the involution:
$
\{H_1,H_2\}=0.
$

Using the ansatz 
\begin{equation}\label{degree2}
\pi'^{ij}=\sum_{k\leq m} a^{ij}_{km}(q_1,q_2)p_kp_m+\sum_{k} b^{ij}_k(q_1,q_2)p_k+c^{ij}(q_1,q_2),
\end{equation}
we find another Poisson bracket with the Poisson tensor 
\begin{equation}\label{p-u}
\pi' = \left( \begin{array}{cccc}
  0 & 0 & \frac{p_2}{2q_1} & 1 \\
  *& 0 & \frac{p_1}{2q_1} -\frac{q_1^2}{6} + \frac{t_1}{2}  - 
  \frac{p_2^2/4 + q_2p_2 - q_2^2 + t_2}{2q_1^2} & -\frac{q_2}{q_1} \\
 * & * & 0 & q_1^2 - t_1 - \frac{p_2^2/4 - 3q_2^2 + t_2}{q_1^2}\\
* & * & * & 0
\end{array}%
\right)\,
\end{equation}
such that it is compatible (see, for instance \cite{fp02}) with the canonical bracket and  $\{H_1, H_2\}'=0$. The eigenvalues of 
the so called recursion operator $\pi'\pi^{-1}$ are roots $u_1, u_2$ of the equation 
\begin{equation}\label{u12}
B(u)\stackrel{def} {=}
=u^2 - \frac{p_2 - 2q_2}{2q_1}u
+ \frac{q_1^2}{6} - \frac{t_1}{2} - \frac{p_1}{2q_1}
+ \frac{\frac{p_2^2}{4} - q_2^2 + t_2}{2q_1^2}=0.
\end{equation} 
Following the papers \cite{g-ts,ts10,ts11,ts12} , we solve the equations 
\begin{equation}\label{ab-poi}
\{B(u),A(v)\}=\frac{B(u)-B(v)}{u-v}\,,\qquad \{A(u),A(v)\}=0
\end{equation}
and find a polynomial  
\begin{equation}\label{a-pol}
A(u)=q_1^2 u + 2q_1q_2 
\end{equation}
of the first dergee. Defining the variables $v_1, v_2$ by the formala 
\begin{equation}\label{v12}
v_{i}=-A(u_{i}),
\end{equation}
we give rise to the transformation 
\begin{align*}
    q_1 =& \sqrt{\frac{v_2 - v_1}{u_1 - u_2}}\,,\quad
    q_2 =\frac{u_1v_2 - u_2v_1}{2\sqrt{(v_2-v_1)(u_1 - u_2)}}\,,\quad
    p_2=\frac{2u_1v_1 - u_1v_2 + u_2v_1 - 2u_2v_2}{\sqrt{(v_2-v_1)(u_1 - u_2)}}
\end{align*}
\[
p_1=\frac{\left(
(u_1 - u_2)(u_1v_1 + u_2v_2)-t_1(v_1-v_2) - t_2(u_1-u_2)-\frac{(v_1 - v_2)^2}{3(u_1 - u_2)}\,
\right)}{\sqrt{(v_2-v_1)(u_1 - u_2)}}
\]
from the original variables $(q_1,q_2,p_1,p_2)$ to the variables $(u_1,u_2,v_1,v_2)$. It can be verified that it
is a canonical tranformation preserving the form of the Poisson bivector \eqref{p-dar}. 
It follows from these formulas and \eqref{darbu} that  the coefficients of the operator $L$  are given by
\begin{equation}\label{UP}
P=\frac{v_2 - v_1}{u_1 - u_2}, \qquad U= \frac{u_1v_2 - u_2v_1}{2 (u_1 - u_2)}.
\end{equation}

 In the variables $u_{1,2}$ and $v_{1,2}$ both Hamiltonians \eqref{I1}, \eqref{I2} become polynomials of the second degree in momenta $v_i$: 
\[
H_1=\frac{v_1^2 - v_2^2}{3(u_1 - u_2)}
-\frac{(u_1^3 - t_1u_1 - t_3)v_1}{u_1 - u_2} + \frac{(u_2^3 - t_1u_2 - t_3)v_2}{u_1 - u_2} + (u_1 + u_2)t_2
\]
and
\[
H_2=\frac{u_1v_2^2 - u_2v_1^2}{3(u_1 - u_2)}
+\frac{u_2(u_1^3 - t_1u_1 - t_3)v_1}{u_1 - u_2}
-\frac{u_1(u_2^3 - t_1u_2 - t_3)v_2}{u_1 - u_2} - \frac{(2u_1u_2 + t_1)t_2}{2}.
\]
To 
 remove the linear terms, we perform the following canonical shift transformation:
\[
w_k=v_{k}-\frac{3}{2}(u_k^3-t_1u_k-t_3)\,,\qquad k=1,2.
\]
In the variables $u_{1,2}$ and $w_{1,2}$  our pair of Hamiltonians can be written as 
$$
H_1=\frac{V_1-V_2}{u_1-u_2}, \qquad H_2=\frac{u_2 V_1-u_1 V_2}{u_2-u_1},
$$
where 
$$
V_{i}=\frac{1}{3} w_i^2-\frac{t_1 t_2}{2} - \frac{3 t_3^2}{4} - 
   \frac{3}{2} t_1 t_3 u_i - \frac{3}{4} t_1^2 u_i^2 + 
   t_2 u_i^2 + \frac{3}{2} t_3 u_i^3 + 
   \frac{3}{2} t_1 u_i^4 - \frac{3}{4} u_i^6.
$$
This means that we are in the frames of the  St\"{a}ckel approach \cite{ts99,fp02}.

Recall that St\"{a}ckel Hamiltonians with $n$ degrees of freedom are characterized by the presence of a family of quadratic integrals of the form 
\begin{equation}\label{hams}
H_k=\sum_{j=1}^n a_{jk}(u_1,\dots,u_n)\left(v_j^2+U_j(u_j)\right),\qquad k=1,\ldots,n\,.
\end{equation}
commuting with each other with respect to the standard canonical bracket \cite{st95}. It is usually
assumed that the matrix ${\cal A}=\{a_{ij}\}$ is non-degenerate. Any of the functions $H_i$ can be chosen as the Hamiltonian in this theory.  The matrix $S={\cal A}^{-1}$   is called the St\"{a}ckel matrix.   The commutativity conditions for $H_i$ are equivalent to the fact that $S_{ik}$ is a function of only $u_k$ for any $i,k$.

For the St\"{a}ckel systems the Hamilton-Jacobi separated relations  have the following form
\begin{equation}\label{st-eq}
{v}_k^2=F_k(u_k)\,,\quad \qquad k=1,\ldots,n\,,
\end{equation}
$$
F(u_k)= U_k(u_k)+\sum_{i=1}^n \alpha_iS_{ik}(u_k)\,
$$
where the constants $\alpha_k$ are the values of integrals $H_i.$ The 
functions $u_i(\tau_1,\ldots,\tau_n,\alpha_1,\ldots,\alpha_n)$ can be found in quadratures from the equations
\begin{equation}\label{jac-inv-g}
\sum_{i=1}^n
\int^{u_i}\frac{S_{ki}(u)\mathrm{d}u}{2\sqrt{F_i(u)}}=\tau_k,
\end{equation}
where $\tau_k$ are times associated with Hamiltonians $H_k$.

In our case, formula \eqref{st-eq} leads to the separation relations
\begin{equation}\label{s-rel}
w_1^2= \mathcal P(u_1)\qquad\mbox{and}\qquad w_2^2= \mathcal P(u_2)\,,
\end{equation}
where
\begin{equation}\label{phi}
\mathcal P(u)=\frac{9}{4}\left(
u^6 - 2t_1u^4 - 2t_3u^3 +\left(t_1^2 - \frac{4t_2}{3}\right)u^2+
+ 2t_1t_3u   + \frac{2t_1t_2}{3} + t_3^2\right)+3\alpha_1 u+3\alpha_2. \end{equation}
Applying formula \eqref{jac-inv-g} and taking into account that the Hamiltonian of the system \eqref{DynSys1}  is $-H_1$, we obtain that the functions $u_{1,2}(x,\alpha_1,\alpha_2)$ are defined by 
\[
\int^{u_1}\frac{\mathrm{d}u}{\sqrt{\mathcal P(u)}}+\int^{u_2}\frac{\mathrm{d}u}{\sqrt{\mathcal P(u)}}=\beta_1, \qquad \quad
\int^{u_1}\frac{u\mathrm{d}u}{\sqrt{\mathcal P(u)}}+\int^{u_2}\frac{u\mathrm{d}u}{\sqrt{\mathcal P(u)}}= \frac{2}{3}x+\beta_2 .
\]
Here $\mathcal P(u)$ is given by \eqref{phi}, $x$ is the independent variable in \eqref{DynSys1} and $\beta_i$ are integration constants.

We are using the traditional form of the inversion of an abelian integral. In the modern form we could rewrite it, using the basis $\displaystyle \frac{\mathrm{d}u}{w},\frac{u\mathrm{d}u}{w}$ of the abelian differentials of the first kind on the curve $w^2=\mathcal P(u)$ and considering the Abel-Jacobi mapping from this curve to its jacobian.

According to \eqref{UP}, the  coefficients of the operator $L$ are given by
$$
P=\frac{w_2-w_1}{u_1-u_2}+\frac{3}{2} (u_1^2+u_1 u_2+u_2^2-t_1), \qquad U=\frac{u_1 w_2-u_2 w_1}{u_1-u_2}+\frac{3}{2} (u_1^2 u_2+u_1 u_2^2+t_3).
$$

Equations \eqref{s-rel}, \eqref{phi} define a hyperelliptic curve of genus two on the $(u,w)$ plane. There is  an alternative  way to introduce a hyperelliptic curve related to our problem, which based on the existence of a common  eigen-function for differential operators $L,$ $M$ and $N$ (see \cite{krich2}).

\begin{proposition} Consider the genus 2 case. Then if the operators $L,$ $M$ and $N$ have a non-trivial common eigen-function 
$$L(\Psi)=\lambda \Psi \qquad M(\Psi)=\mu \Psi, \qquad N(\Psi)=\nu \Psi$$
then, up to a translation  $\lambda\mapsto \lambda + c_1$\footnote{We shift a singular point on the curve to $\lambda=\mu=0$.},
\begin{equation}\label{curve}
\mu^3=\lambda^4  +  t_1 \lambda^2 \mu +  \frac{t_2}{3} \mu^2+ t_3 \lambda^3  -  \frac{H_1}{3} \lambda \mu - \frac{1}{6} (2 H_2+t_1 t_2) \lambda^2.
\end{equation}
\end{proposition}
\begin{lemma}
Any algebraic curve of the form
\begin{equation}\label{curva}
 \mu^3=\lambda^4 +  z_1 \lambda^2 \mu + z_2 \lambda^3 + z_3 \mu^2 + 
 z_4 \lambda \mu + z_5 \lambda^2 
\end{equation}
can be mapped  to the hyperelliptic form
\begin{equation}\label{trans}
\eta^2 = \xi^6 - 2 \xi^4 z_1 - 2 \xi^3 z_2 + \xi^2 (z_1^2 - 4 z_3) + 
   \xi (2 z_1 z_2 - 4 z_4) + (z_2^2 - 4 z_5)
\end{equation}
by the transformation
\begin{equation}\label{trans2}
\lambda = \frac{1}{2} (\eta+\xi^3 - \xi z_1 - z_2), \qquad  \mu =  \frac{1}{2} \xi \,(\eta + \xi^3 - \xi z_1 - z_2).
\end{equation}
The inverse transformation has the form
\begin{equation}\label{inverttrans}
\xi =\frac{\mu}{\lambda},\qquad  \eta =  - \frac{\mu^3}{\lambda^3} + 2 \lambda + \frac{\mu z_1}{\lambda} + z_2.
\end{equation}
\end{lemma}

For \eqref{curve} we obtain
\[
\eta^2=\xi^6-2 t_1 \xi^4-2 t_3 \xi^3+\left(t_1^2-\frac{4}{3} t_2\right)\xi^2+\left(\frac{4}{3} H_1+2 t_1 t_3\right) \xi+\frac{4}{3} H_2+\frac{2}{3} t_1 t_2+t_3^2\,,
\]
which coincides with the obtained curve \eqref{s-rel}-\eqref{phi} after the change of coordinates 
\[\xi = u\,\qquad \eta = \frac{2}{3}\,w\,. \]

\section{Conclusion}
 
In this paper we have considered a particular system of the Novikov equations, concentrating on its dynamic aspects; the algebro-geometric ones will be considered in the subsequent publications.

The general Novikov equations of rank 1 correspond to such commutative subrings of the ring of differential operators that the orders of the operators of the subring assume almost all the orders. The set of orders of these operators is closed under addition and hence the Novikov equations of rank 1 are parametrized by such submonoids \rm $O\subseteq\mathbb{N}$ that
$
O+O=O
$
and
the complement $\mathbb{N}\setminus O$ is finite. (The theory of such submonoids is quite non-trivial and dates back to the 19-th century, see \cite{sylv} and consult, e.g. \cite{ust} for the current state).

The KdV hierarchy corresponds to the monoids generated by 2 and an odd number;
the present paper deals with the one generated by 3,4,5. 
In the future publications we hope to develop the theory of Novikov equations corresponding to the general monoids $O$ and, in particular, 
to apply it to the theory of special divisors on algebraic curves.  

\subsection*{Acknowledgements}

The authors are grateful to  V.Buchstaber and A. Zheglov  for useful discussions. 
The research of VVS is supported by the MSHE ``Priority 2030'' strategic academic leadership program.
The reasearch of AVT was carried out with the financial support of the MSHE in the framework of a scientific project under agreement No. 075-15-2024-631.

\end{document}